\documentclass[sigconf,preprint]{acmart}
\pdfoutput=1
\makeatletter                   
\def\mdseries@tt{m}             
\makeatother                    
\usepackage[plain]{fancyref}
\usepackage[draft=true]{minted} 
\usepackage{color}
\usepackage{hyperref}           
\hypersetup{
    colorlinks=true,
    linkcolor=blue,
    filecolor=red,      
    urlcolor=magenta,
    breaklinks=true,            
}
\usepackage{breakurl}           
\begin{document}
\sloppy                         

\title{Designing a Future Worth Wanting: Applying Virtue Ethics to HCI}

\author{Tim Gorichanaz}
\email{gorichanaz@drexel.edu}
\orcid{0000-0003-0226-7799}
\affiliation{%
  \institution{Drexel University}
  \streetaddress{3675 Market Street}
  \city{Philadelphia}
  \state{Pennsylvania}
  \country{USA}
  \postcode{19104}
}

\begin{abstract}

Out of the three major approaches to ethics, virtue ethics is uniquely well suited as a moral guide in the digital age, given the pace of sociotechnical change and the complexity of society. Virtue ethics focuses on the traits, situations and actions of moral agents, rather than on rules (as in deontology) or outcomes (consequentialism). Even as interest in ethics has grown within HCI, there has been little engagement with virtue ethics. To address this lacuna and demonstrate further opportunities for ethical design, this paper provides an overview of virtue ethics for application in HCI. It reviews existing HCI work engaging with virtue ethics, provides a primer on virtue ethics to correct widespread misapprehensions within HCI, and presents a deductive literature review illustrating how existing lines of HCI research resonate with the practices of virtue cultivation, paving the way for further work in virtue-oriented design.

\end{abstract}





\maketitle

\begin{quote}
	They constantly try to escape \\
	From the darkness outside and within \\
	By dreaming of systems so perfect that no one will need to be good.
	
	\medskip
	
	\hspace{1cm} ---T. S. Eliot, ``Choruses from \emph{The Rock},'' 1934 
\end{quote}

\section{Introduction}

The story of technology is one of continually growing power, reach, speed and scale. Networked computing technologies enable us to affect each other more than ever before---and to do so more quickly and significantly. This trajectory has inspired countless developments, from real-time video calling worldwide to rapid vaccine development and distribution. But it has also brought along difficult sociopolitical consequences and moral quandaries, from the spread of misinformation \cite{Piccolo2021} to the difficult economics of gig work \cite{Alkhatib2017} and on and on; and these issues present challenges for human wellbeing and flourishing in the digital age. 

Amidst this trajectory, the moral side of technology is becoming impossible to ignore. While technology design is often oriented toward commercial values such as productivity and efficiency \cite{Hallnas2001,Morozov2013,Walker2011}, scholars in human--computer interaction (HCI) and allied fields have for decades been researching and developing methods for orienting technology design toward essential human values such as justice, reflection and meaning \cite[e.g.,][]{Brey2015,Desmet2012,Friedman2002,Mekler2019}. Of late, such discussions have entered the public discourse as well \cite{Gjika2019,Ouchchy2020,Shipman2020}.

HCI researchers and designers, as well as users and other stakeholders, seem to be searching for ethical guidance for the digital age. In all fields of technology, discussions of ethics are on the rise. Commonplace are (often implicit) burning questions such as what makes life worth living and how we can make the world a better place through design. Ethics-based methods for design are proliferating \cite{Chivukula2021}, and in universities there is a growing interest in teaching ethical approaches to sociotechnical design and evaluation \cite{Fiesler2020}. The past few years have seen the inception of agencies dealing with such issues, including the Center for Humane Technology (founded in 2018) and the Sacred Design Lab (2020), as well as countless academic research centers, including the Technology Ethics Center at the University of Notre Dame (2019) and the Ethics, Technology, and Human Interaction Center at Georgia Tech (2020)---to say nothing of the myriad such initiatives that existed prior. 

To facilitate these efforts, continued translational work is needed to incorporate developments from moral philosophy into computing and vice versa \cite{Floridi2013,Su2019}. In HCI, the majority of work in this vein has focused on two of the three major ethical traditions: deontology, which focuses on rules of behavior; and consequentialism, which focuses on outcomes of behavior. There has been a noticeable lack of meaningful engagement with the third major ethical tradition, virtue ethics. This approach to ethics focuses on the traits, situations, motivations and actions of moral agents rather than on the outcomes of actions or on rules governing action \cite{Rachels2015}. Virtue ethics explains how moral qualities are identified, how character is developed, and how moral dilemmas are negotiated in real life. Over the past decade, a number of scholars have made the case that virtue-based approaches to design are uniquely well suited to today's sociotechnical environment \cite[e.g.,][]{Heersmink2018,Vallor2016}, but HCI has not yet integrated these arguments. 

This paper provides an overview of virtue ethics for application in interaction design. While typical work in HCI, particularly that presented at conferences, has focused on practical application rather than reflective and theoretical discussions, this paper seeks to provide a bridge between the theoretical and the applied as part of an ongoing move toward deeper engagement with theoretical issues in HCI. This paper begins by briefly defining virtue ethics, reviewing existing work in HCI that engages with virtue ethics, and outlining an argument for adopting virtue ethics over other ethical theories in design. Then, with an eye toward demonstrating the applicability of virtue ethics to HCI, the paper discusses the major tenets of virtue ethics, focusing particularly on the practices for cultivating virtue. The paper illustrates how existing work in HCI resonates with those practices, leading to opportunities for direct engagement with virtue ethics based on existing lines of research in HCI.

\section{An Overview of Virtue Ethics}

To put it telegraphically, ethics is the study of the good life. Broadly, there are three main approaches to ethics---deontology, consequentialism, and virtue ethics---each of which separates good from bad actions by focusing on a different aspect of action: deontology focuses on rules and duties that guide action, consequentialism focuses on outcomes of actions, and virtue ethics focuses on certain qualities of the person doing the action \cite{Rachels2015}.

Virtue ethics posits that people can achieve the good life through cultivating specific moral qualities, called virtues, within themselves \cite{Hursthouse2018}. What qualities are considered virtues is socially local, rather than universal; still, some virtues are upheld across cultures and epochs, such as courage, justice and honesty \cite{MacIntyre1981}. Besides helping identify the virtues, virtue ethics contends that we can become more virtuous. That is, unlike certain human qualities such as height, the virtues are not inborn or fixed; rather, we must practice and grow them. Moreover, the virtues do not exist in isolation. Virtue ethicists speak of a person's \emph{character}, or their virtue writ large, beyond simply observing that a person exhibits this or that particular virtue \cite{Homiak2019}. Thus, according to most theorists, the virtues are united or integrated through a person's practical wisdom, i.e., the ability to act discerningly given the particulars of each situation \cite{Barry2017}. For example, a would-be courageous person who is lacking in practical wisdom may, in the end, simply be foolhardy \cite{AristotleEthics}. Additionally, virtue is not just about action, but also perception, motivation and justification; being virtuous is about doing the right thing for the right reasons \cite{Wright2021}.  

As mentioned, crucially, the virtues are not fixed traits we are born with; rather, they are practiced and honed throughout our life. How is this done? A recent and highly regarded account of this is given by Shannon Vallor in \emph{Technology and the Virtues} \cite{Vallor2016}. In this book, Vallor synthesizes a tremendous literature spanning the philosophy of technology and several other philosophical traditions, as well as case studies of several emerging technologies. In particular, she analyzes three distinct traditions of virtue ethics---Aristotelianism from Greece, Confucianism from China, and Buddhism from India---with the aim of developing the foundation of a global morality for the 21st century and beyond. As part of this work, she provides a framework of seven practices by which the virtues are cultivated: 

\begin{description}
	\item [Moral Habituation] Learning by doing, not just theorizing, and improving gradually over time
	\item [Relational Understanding] Recognizing our roles and relationships with others
	\item [Reflective Self-Examination] Checking in on ourselves with respect to moral aspirations
	\item [Self-Direction] Choosing our goals and taking steps toward those goals
	\item [Moral Attention] Becoming aware of and tending to morally salient facts in the environment 
	\item [Prudential Judgment] Choosing well among available options 
	\item [Extension of Moral Concern] Doing good for others as well as ourselves 
\end{description}

Just like the virtues themselves, these practices are mutually reinforcing. Moreover, they are not just steps to unlock virtues, but rather they are ``constitutive and enduring elements of virtue itself'' \cite[p.~66]{Vallor2016}. Each practice will be discussed in further detail below, in connection with relevant existing work in design. But first, the following section presents an argument for adopting virtue ethics as a guiding framework in interaction design.

\section{Why Virtue Ethics in HCI}

Virtue ethics is an approach to ethics with roots in ancient traditions from several cultures dating back millennia \cite{Vallor2016}. Over the last few centuries, it gradually fell out of favor, giving way to deontology (which resonated with the Judeo-Christian religions' focus on laws) and later consequentialism (growing out of the Enlightenment ideals of calculated rationality and universality) \cite{MacIntyre1981,Rachels2015}. But over the past half-century, there has been a renewed interest in and development of virtue ethics \cite{MacIntyre1981,Rachels2015,Vallor2016}. This resurgence has yet to meaningfully penetrate HCI, wherein deontology and consequentialism are still the most broadly understood and applied ethical theories \cite{Chivukula2021,Zoshak2021}. Yet there are good reasons for HCI to engage with virtue ethics, which are discussed in this section.

Let us begin with a reflection on our sociotechnical situation. With today's technologies, humanity has unprecedented power, including the ability to annihilate our own existence \cite{Light2017}. Our efforts have affected our lived environment, as manifest in the changing climate. Our institutions are capable of producing new sociotechnical platforms that drastically change our way of life in a matter of years (e.g., smartphones, social media). Even individuals have further reach than ever before. These new technologies, which proliferate through the potency of network effects, introduce possibilities and dangers whose consequences are impossible to predict (genetic modification with CRISPR/Cas9, proposals for geo-engineering to curb climate change, social media policies intersecting with global politics, etc.). Moreover, our environment is also becoming less predictable (fires, droughts, extreme storms). Given all this, the future is less predictable than ever before \cite{Allenby2011,Vallor2016}. 

How does this bear on morality? Recall that ethics is the search for the good life. Life, of course, depends on the material qualities of our lived environment as well as our technologies. Most ethical traditions rely on the moral landscape of the future being more or less as it is in the present, with technological change coming slowly enough for our moral faculties to evolve alongside it \cite{Vallor2016}. Mill, for example, in his introduction to utilitarianism, wrote that the theory would work ``as long as foresight is a human quality'' \cite[p.~25]{Mill2001}.

In this light, we can appreciate reasons deontology and consequentialism fall short as moral guides today---many of which have already been noted by HCI scholars \cite{Friedman2002,Light2017,Millard2019,Zoshak2021}. Deontology seeks to provide rules for moral action. It is, of course, impossible to list an exhaustive set of rules, and situations arise where rules conflict. In any case, with such an opaque future, we cannot know ahead of time what the moral rules should be \cite{Vallor2016}. Even systems with one or a few rules, such as Kant's categorical imperative (i.e., an action is good if we would want everyone to do it), fail us, as they require us to know relevant facts about life in the future while assessing each rule. For example, consider asking yourself in 2005 whether it would be good for everyone to be on Facebook. Consequentialism, on the other hand, does not rely on specified rules, but it does require that we be able to know and predict the consequences of our actions. At its worst, this lends itself to the perfectionism of endlessly seeking to optimize outcomes \cite{Vallor2016}. But moreover, given the vicissitudes of emerging technologies, we certainly cannot predict the long-term consequences of our actions; and even in the short term, actions may have far-flung knock-on effects due to the affordances of networked digital technologies \cite{Jasanoff2016}. 

In our sociotechnical environment, we cannot rely on fixed rules of conduct (we need to know when to bend the rules or even rewrite them) or on predicting outcomes. Virtue ethics, on the other hand, provides a balanced, dynamic and responsive framework for discerning and moving toward the good life using tools and concepts that have been with humanity for millennia and which are still relevant, even given the tremendous change of the last few centuries \cite{Heersmink2018,Vallor2016}. Vallor writes: 

\begin{quote}
Virtue ethics is a uniquely attractive candidate for framing many of the broader normative implications of emerging technologies\ldots~Virtue ethics is ideally suited for adaptation to the open-ended and varied encounters with particular technologies that will shape the human condition in this and coming centuries. Virtue ethical traditions privilege the spontaneity and flexibility of practical wisdom over rigid adherence to fixed rules of conduct---a great advantage for those confronting complex, novel, and constantly evolving moral challenges such as those generated by the disruptive effects of new technologies. \cite[p.~33]{Vallor2016}
\end{quote}

\section{Prior Work on Virtue Ethics in HCI}

Though deontology and consequentialism are the most frequently applied ethical theories within HCI, there are some precedents for applying virtue ethics. While this paper is not meant to provide a systematic, comprehensive literature review of the subject, an illustrative presentation of this work is in order. To explore the existing work in HCI that deals with virtue ethics, a literature review was conducted in July 2021. This review was targeted toward works that overtly mentioned or employed virtue ethics. This began with a full-text search in the ACM Digital Library for ``virtue ethics'' in SIGCHI-sponsored venues, returning 17 results. To supplement this, searches were conducted for texts containing ``virtue ethics'' and ``hci'' in Google Scholar, IEEE Xplore and LISTA (Library and Information Science and Technology Abstracts). For relevant articles, citations were traced forward and backward using Google Scholar. While no date limitations were used in any of these searches, it is striking that none of the resulting articles was from earlier than 2001, and the vast majority were from 2019 onwards. 

For the purposes of discussing the body of literature retrieved, a thematic analysis was conducted. Four themes characterize this literature: human--robot interaction; HCI methodology; design education; and the design process. In the field of human--robot interaction, virtue ethics has become a growing subject of discussion over the past few years. Some of this work explores the question of robots having moral worth as beings \cite{Coeckelbergh2021,Sparrow2016,Zoshak2021}, and other literature explores whether robots could be competent moral actors \cite{Kim2021,Poulsen2018,Wen2021,Williams2020}. Whereas most of this work discusses robots in general, some of authors have focused on automated vehicles in particular \cite{Gerdes2020,Thornton2017}. Next, some scholars have applied virtue ethics to HCI research methodology, including discussing the ethical dimensions of visualization research \cite{Correll2019} and consent in human subjects research \cite{Strengers2021}. Third, virtue ethics has also entered the conversation regarding design education. Virtue ethics has formed the philosophical grounding for classroom activities in design fiction \cite{Baumer2018}, data science \cite{Shapiro2020}, and system analysis \cite{DiPaola2020}. In a broader contribution, Pierrakos et al. argue that ethics education for engineers should move from an emphasis on rule-following to one of moral character, i.e., virtue \cite{Pierrakos2019}. 

Finally, the bulk of the literature retrieved relates to the conceptual foundations of the design process, which is most relevant to the present paper. Some of this work simply mentions that deontology and consequentialism are the norm in design ethics but acknowledges virtue as a fruitful path to consider \cite{Chivukula2021,Friedman2002,Light2017,Millard2019}. (This is also mentioned among the works in human--robot interaction \cite{Gerdes2020,Williams2020,Zoshak2021}.) Light et al. specifically mention virtue ethics as a useful framework given the dynamic, intersecting crises that humanity is facing at present \cite{Light2017}. Next, a few contributions provide guidance for applying virtue ethics, including frameworks for virtue-based engineering \cite{Barford2019,Moriarty2001} and interaction design \cite{Barry2017}. Some authors have enumerated particular virtues to be used in design \cite{Kirkham2020,Steen2013}, and one contribution discusses how design fiction can offer a path for reflecting on virtues in design \cite{Berner2019}. More granularly, Chivukula et al. \cite{Chivukula2021} provide a listing of eight particular design techniques using virtue ethics as part of their survey of ethical design techniques. However, only two of these are explicitly rooted in virtue ethics; four others resonate with but do not explicitly mention virtue ethics; and two (Borning et al. and Zhou) seem better described as deontological or consequential rather than virtue-oriented.

Much of the work reviewed, unfortunately, suggests misunderstandings of virtue ethics. For example, van Berkel et al. write, ``The implementation of virtues in AI applications is inherently challenging, e.g. how would one go about quantifying a courageous algorithm through a rule-based approach?'' \cite{vanBerkel2020}. This seems to confuse virtue ethics with deontology, as virtue-oriented approaches are categorically \emph{not} rule-based; moreover, virtue ethicists would not suggest that we attempt to creat a ``courageous algorithm,'' but rather that we help \emph{people} practice courage. Besides such fundamental errors, others have characterized virtue ethics as focused only on the solipsistic development of one's character traits, overlooking the sociality, context-sensitivity and dynamism of virtue ethics. For example, Chivukula et al. write that ``traditional systems of professional ethics~\ldots~often fixate on consequentialist, deontological, or virtue ethics [rather than] a pragmatist ethics that values both the designer and her character \emph{and} the unique complexity of the design situation'' \cite[p.~9]{Chivukula2019}. But virtue ethics, properly understood, \emph{does} value the complexity of the situation as well as the actor's character. More recently, Chivukula et al. have defined virtue ethics in design narrowly as focusing on the designer's character alone. Again, this overlooks the way virtue is embedded in moral communities, not just individualistic actors. Such a perspective then misses the possibility that designers could create products that \emph{help users cultivate virtues}, not just that designers can cultivate their own virtues. 

All in all, the review of prior work suggests that the field of HCI is in need of a more accurate and complete understanding of virtue ethics, including how it is eminently compatible with emerging ethical theories under discussion such as pragmatic ethics and care ethics, as well as more guidance for applying virtue ethics to existing frameworks. The previous sections may have already gone some way toward helping the field achieve the former; the next section concerns the latter. 

\begin{figure}[bt]
 \centering
 \includegraphics[width=\columnwidth]{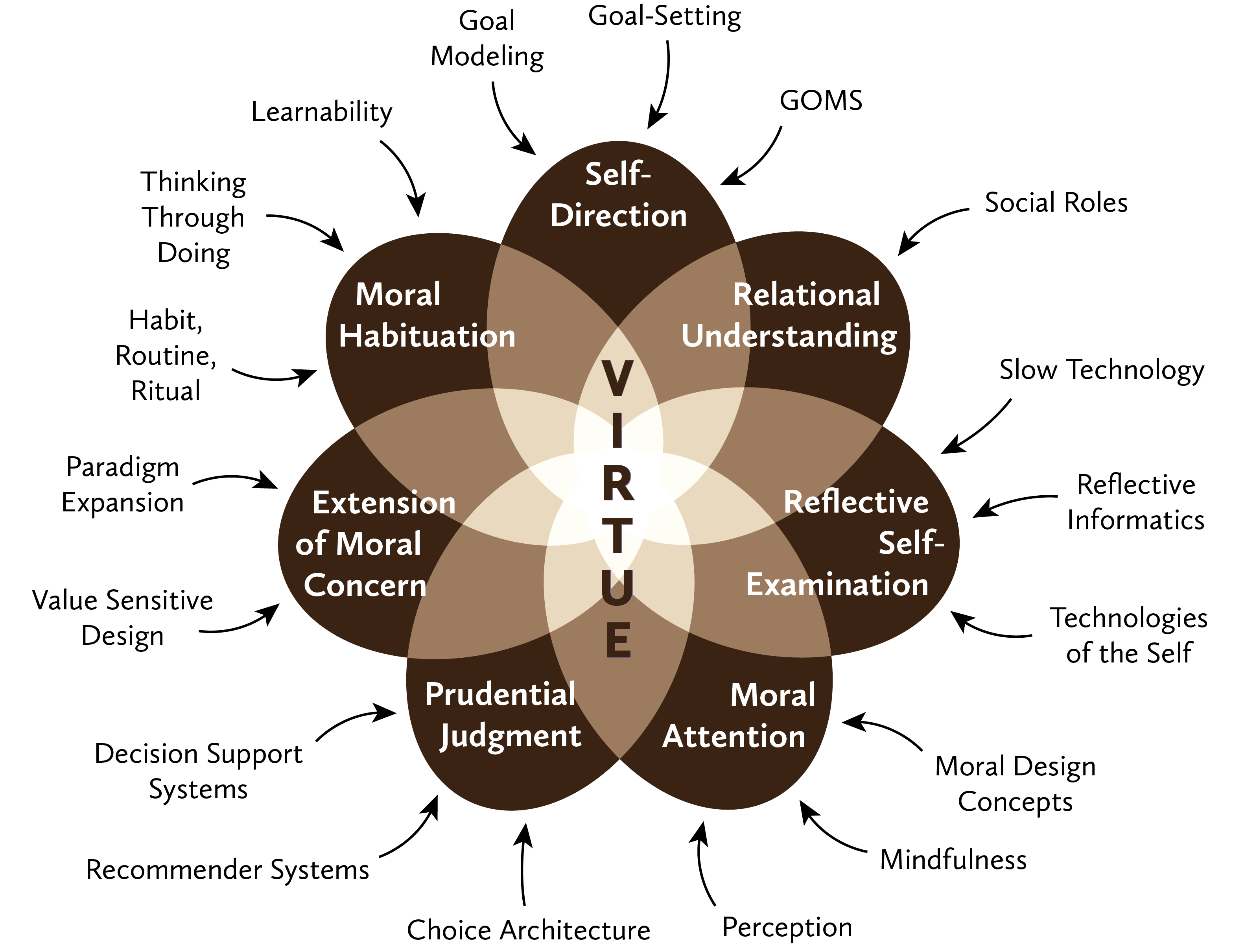}
 \caption{How the seven practices of virtue cultivation \cite{Vallor2016} (inner text) connect with existing literature in HCI (outer text)}
 \Description{Diagram showing how existing areas of HCI research, such as goal-setting, reflective informatics and choice architecture, connect with virtue ethics by way of contributing to practices for cultivating the virtues.}
 \label{fig:flower}
\end{figure}

\section{Connecting the Practices of Virtue Cultivation to Existing Research in HCI}

One aim of this paper is to demonstrate how virtue ethics can be applied to HCI research and design. To that end, this section identifies connections that already exist implicitly between HCI and virtue ethics by using the seven practices of virtue cultivation outlined above as an analytical framework. Through this framework, it is possible to discern and classify existing work in HCI that may readily contribute to virtue-oriented design. This is summarized in Figure~\ref{fig:flower}. 

This section serves to help designers and researchers in many areas of HCI to better appreciate how their work already may contribute to virtue cultivation. In many cases, designing to support morality requires only a slight shift of focus in existing research programs. Moreover, others interested in moral design can recognize and continue to organize the wealth of relevant existing work in HCI and build upon this strong but as yet disparate foundation. 

The topics discussed in this section arose through years of research and teaching a range of topics in HCI and design as well as open-ended literature searching. This section does not purport to be exhaustive or systematic. Rather, it offers examples and illustrations of how virtue ethics may be applied in range of topics across HCI and design. The heuristic value of this section lies in drawing connections and pointing to opportunities in research and design.

The following subsections are dedicated to the seven practices of virtue cultivation described by Vallor \cite{Vallor2016}. In each section, the practice is first defined in the context of virtue ethics. Next, connections to existing HCI literature are described for illustration. Finally, opportunities are pointed out for further research to take the existing work in a more virtue-oriented direction. 

\subsection{Moral Habituation}

According to virtue ethicists, we become virtuous primarily by acting morally, not by thinking or reading about morality \cite{Vallor2016}. Moral habituation implies gradual improvement over time as we strive to do the right thing in each situation that confronts us. It also involves justification; true moral habituation is not just rote doing, but also understanding the reasons for that doing. With moral habituation, we have a reason for acting, the action is valued as good, and we repeat that action, which shapes our ongoing emotional states. This progress leads us to experience joy in performing moral action, which in turn further strengthens our commitment to doing good \cite{Vallor2016}.

Moral habituation relates to several threads of HCI literature. The notion of learning through action recalls the maxim that professionals (not least designers) think through doing, which Donald Sch\"on described as reflection-in-action \cite{Schon1984}. Users, too, learn through doing, which is encapsulated in the concept of learnability, a facet of usability \cite{Nielsen2012}. As well, there is a rich literature in HCI on users' repeated use of systems in various contexts, framed with concepts such as habit \cite{Chow2019,Lange2015,Lankton2012,Pinder2018}, routine \cite{Begole2002,Pentland2008,Su2013,Tolmie2002}, and ritual \cite{Baharin2016,McVeighSchultz2018,Ozenc2018,Petrelli2014}. 

However, this prior work has conceptualized habituation amorally---that is, it does not entail an orientation toward right action. Still, the work on habit, routine and ritual in HCI could easily be a site for applied virtue ethics if the focus turns toward the morally good and not just the desirable or profitable. 

\subsection{Relational Understanding}

Virtue ethics conceptualizes the human person as a fundamentally social being---one formed through a network of relationships. Vallor writes, ``In contrast, both Kantian and utilitarian ethics portray the self as an independent rational agent who can and ought to act \emph{autonomously}, without relying on the external guidance of others'' \cite[p.~76]{Vallor2016}. The practice of relational understanding is, first, the continual pursuit of a more precise and accurate view of our social relationships; and second, building the skill of using this understanding in prudential judgments in response to the moral exigencies of each particular relationship in each particular circumstance \cite{Vallor2016}. 

In HCI, there has been a sizable literature examining human social roles in various sociotechnical contexts. Previous work has examined social roles in family life \cite{Ammari2016,Balaam2013,Kimani2010}, the workplace \cite{Barlow2013} and online support communities \cite{Yang2019}, and researchers have investigated gender roles regarding emerging technologies \cite{Pater2019,Stavrositu2008,Yassien2021}. Research in this area has been marshaled toward the automated recognition of social roles, for example in the workplace \cite{Anderson2016,Chandra2018,KremerDavidson2017,Wilson2011}, in teaching \cite{Howley2014}, and across contexts \cite{Anderson2016,Anderson2019,Papapetrou2014}. Additionally, work has been applied to the development of automated systems such as chatbots \cite{Stergiou2001,vanHooijdonk2021} and social robots \cite{Grace2017,Howley2014,Huber2016}.

Just as with prior work related to moral habituation, the research surveyed in this section has been done without any particular moral intent. The existing work could be deepened by foregrounding the particularly moral dimensions of roles in sociotechnical contexts, and future work can be channeled toward virtue-based design. For example, technologies that can discover social roles in context could show these findings to the user, offering them an opportunity for self-reflection and allowing them to consider the moral relevance of these roles (e.g., what duties and opportunities a given relationship might suggest) and to adopt new roles as prudent.

\subsection{Reflective Self-Examination}

Third, virtue ethics contends that moral principles are not one-size-fits-all. Rather, any moral decision must take into account our personal strengths and weaknesses. Assessing these accurately requires a habit of reflective self-examination. The goal of such examination is to discern how well we are conforming to the ideal moral self we aspire to be---taking into account our beliefs, thoughts, feelings and actions. Reflective self-examination engenders within us a sense of responsibility to correct our shortcomings and stirs up a sense of joy in doing so \cite{Vallor2016}. 

The practice of reflective self-examination resonates with philosophical discussions of technologies of the self. Foucault, who originated the term, describes for example how self-care was enacted in Ancient Greece and Rome through sociotechnical practices such as keeping personal notebooks and writing letters \cite{Foucault1988}. In HCI, related work is broadly part of the Slow Technology movement \cite{GrosseHering2013,Hallnas2001}. Within this movement, many have examined how technology can aid people in self-reflection \cite{Baumer2015,Fleck2010,Gibson2017,Mols2016,Odom2019}. HCI scholars have shown how technology can support self-reflection in several ways: by providing reasons to reflect \cite{Fleck2010,Saksono2017}; by providing the conditions for reflection \cite{Baumer2015,Mols2016,Saksono2017}; by supporting reflection at different levels of depth \cite{Fleck2010}; and by guiding people through the stages of reflection \cite{Mols2016,Saksono2017}.

In what is now a familiar refrain, most of this work has proceeded without reference to an external set of values. In this literature, reflection is taken to be a personal matter, for instance a path to greater pleasure or self-improvement. However, some recent work has tied reflection explicitly to values and ethics, such as exploring how reflection on our values may aid in combating online misinformation \cite{Arif2018}. Future work could explore how technology can encourage and assist people in reflecting specifically on their moral strengths and weaknesses, their virtuous development, etc. 

\subsection{Self-Direction}

As humans, our actions are generally goal-directed. These goals may be big or small, long-term or short-term, implicit or explicit. However, we seldom consider why we have particular goals or wonder if we should have different ones, especially in the moral realm. But ethics is about intention and justification as much as action; and virtue ethics in particular asks us to generate a vision of our ideal future moral self---as Vallor writes, ``the type of being that is \emph{worthy of my becoming in this world}'' \cite[p.~97]{Vallor2016}---to provide direction to our efforts. Without such insight, all the moral intention and action in the world may be fruitlessly misdirected. Indeed, virtuous development is said to be motivated in the first place by our observing \emph{moral exemplars} in the community---those people we admire for their morality and thus seek to emulate \cite[see also][]{Zagzebski2017}. The practice of self-direction entails courageous movement down the path we have chosen for our life toward realizing our ideal future self. Because this is difficult, Vallor says that we must follow this path not for social gain but out of a true desire to cultivate righteousness for its own sake \cite{Vallor2016}.

Work related to goals has a long history in HCI. An early and still applicable theoretical model in HCI is GOMS (goals, operations, methods and selection rules), which posits that people select and carrying out particular methods meant to achieve their goals and provides methods for evaluating and optimizing systems with respect to users' goals \cite{Card1983}. A sizable literature engages with goal-setting theory from psychology to examine how technology can support goal achievement; this work has been applied to areas such as education \cite{Basavaraj2018,Magyar2020,Peters2017}, gamification \cite{Tondello2018}, fitness tracking \cite{Konstanti2018,Munson2012,Niess2018}, energy consumption \cite{Scott2011}, and overcoming challenges related to disability \cite{Geurts2019,McNaney2018}. Another line of HCI research has sought to develop automated methods for discerning and modeling users' goals \cite{Alslaity2021,Smith2010,Zhu2012} to create adaptive user interfaces that help people achieve their goals \cite{Faaborg2006,Lawrance2010,Woodruff1998}. In this vein, recent work has explored how AI systems can suggest goals to users and allow users to reflect on those goals \cite{Mitchell2021}.

HCI research on goals has contributed to our understanding of how people's goals form and how goal setting and achievement can be assisted with technology. From the perspective of virtue ethics, however, this work has some limitations. The literature on GOMS, for instance, has taken people's goals at face value, seeking to help people achieve their goals rather than helping them choose from among a set of goals. GOMS is also tuned toward quantitative evaluation of usability in specialized settings, rather than broader questions of user experience and morality ``in the wild.'' The work related to goal-setting theory and goal modeling does attend to selecting and defining goals, but this has been done with respect to goal specificity and difficulty, not any moral dimensions. Further research applying virtue ethics to goal-setting and goal-oriented user interfaces may be able to use similar methods to assess how to support people in identifying and achieving goals related to the cultivation of virtues; moreover, it may result that there are additional dimensions of goals relevant to morality that should be considered, besides simply difficulty and specificity. As well, future research in this area may examine the identification of moral exemplars among designers and users, as these are at the root of self-direction. 

\subsection{Moral Attention}

Next, becoming virtuous requires that we understand when moral action is called for. To do so, we must sensitize ourselves to the features around us that are morally relevant, which Vallor calls moral attention. At our best, we will be able to say not only that some feature is morally salient, but also why it is and what to do about it. Vallor offers the simple example of a hungry coyote stalking toward a defenseless toddler. Observing this, any of us would understand the danger for the child and be stirred into action \cite{Vallor2016}. But most situations are not so clear-cut. If we encounter a person on the street studying a map, we read their body language and facial expressions to determine if they need help, or if they're just planning. If we see a bedraggled person watching us from a city bench, do we avert our eyes, cross the street, offer them some spare cash, bring them a new set of clothes, or invite them into our home for dinner and a wash? There are any number of subtle features of the situation that help us make this decision. Indeed we may make a decision that upon reflection we realize was the wrong one---hence the need for reflective self-examination. More difficult still are those encounters and situations that are protracted and dynamic: how best to act during an emerging novel disease pandemic, for example? Cultivating moral attention gradually helps us discern what each particular moral situation calls us to do; sometimes this involves adapting or flouting social conventions or normative rules that otherwise would apply. In this way, we can respond flexibly to changing situations. 

In HCI, there is a long tradition of research on human perception, particularly for the design of effective displays \cite{Wickens2004}, and on attention, especially in complex sociotechnical environments amidst the threat of interruption \cite{Bakker2016,Roda2011}. This work tends to be pedestrian, oriented toward making sure technologies are well-suited to the human organism. For example, a large literature is dedicated to human attention and distraction while driving to increase road safety \cite[e.g.,][]{Schneiders2020}. But the practice of moral attention calls for a deeper sense of perception and attention, one that integrates human experience amidst a dynamic environment with moral interpretation. There has been some work directly in this vein, looking for example at students' and engineers' ethical awareness in decision making \cite{Boyd2021,Gray2019b}. There is also a vein of work in HCI on cultivating mindfulness, which is generally defined as a form of attentional training \cite{Akama2015,Dauden2018,Sas2015,Sliwinski2015,Thieme2013}. There has also been work in HCI on the moral features of our digital environments, equipping us with concepts such as dark patterns \cite{Gray2018,Mathur2021}, ghost work \cite{Gray2019}, etc.; having such concepts is a vital step in training our moral attention. 

While much of the work in HCI on attention and perception is not focused on morality, there is an emerging literature that has the capacity to connect to moral attention. Existing work on human attention, distraction, interruption, and so on can be channeled to create technologies that do not completely consume people's attention, allowing them autonomy and flexibility in where they attend to and therefore to cultivate their moral attention. Future work may continue to identify ethical concepts and moral dimensions of existing user interfaces and sociotechnical infrastructure, foster public discussion around them, and equip people with moral actions they can take in response. 

\subsection{Prudential Judgment}

Central to ethics is the question of making decisions and taking action, and likewise Vallor argues that this list of practices would be incomplete without prudential judgment. This is ``the ability to choose well, in particular situations, among the most appropriate and effective means available for achieving a noble or good end'' \cite[p.~105]{Vallor2016}, and having such an ability is necessary for acting in unanticipated or rapidly evolving situations. Prudential judgment involves interpreting our circumstances and emotions correctly, understanding our options and their possible consequences, and choosing the best among them.

One way in which prudential judgment could be applied to HCI is in the design of interactive systems to help users make better moral judgments. Indeed, since their inception, computers have been used to support human decision-making, whether by providing data, calculations or models. An early theoretical vision for this was J. C. R. Licklider's 1960 description of ``man--computer symbiosis'' that would enable people to make better-informed decisions more quickly and easily \cite{Licklider1960}. As such, for the entire history of HCI, researchers have contributed to the development of decision support systems, most notably in business management \cite{Blanning1979,Meador1973}. Today, the development of decision support systems still represents a major line of work in HCI, particularly in healthcare \cite{Rundo2020,Yun2021}. More recently, this sort of work has been conceptualized as \emph{choice architecture} \cite{Jameson2014}, which is a matter of ``organizing the context in which people make decisions'' \cite[p.~3]{Thaler2008}, often including subtly nudging users toward desirable choices. HCI researchers in this area have sought to operationalize the psychology of human decision-making by creating technologies to help people make judgments \cite{Jameson2014}. This research area is still nascent, but ongoing research provides methods to support designers in choice architecture \cite{Caraban2019,Caraban2020,Jameson2014} and explores how these ideas could be integrated into AI recommender systems \cite{Jesse2021}.

Some may criticize nudging as inherently immoral on the grounds that it works through underhanded manipulation. However, there are other ways to accomplish nudging, and the majority of nudges in the HCI literature stimulate conscious reflection (rather than unconscious response) through transparent (rather than black-box) mechanisms \cite{Caraban2019}. Such systems may be more morally defensible, particularly when they nudge us to carry out actions that we have already committed to wanting to carry out, such as commuting by bike more often than car. While decisions around self-improvement and productivity may be morally relevant, most work on nudging in HCI is not explicitly oriented toward moral outcomes. Future work in HCI could move in that direction. Indeed, there is a growing literature outside of HCI on the ethics of nudging \cite[e.g.,][]{Bovens2009,Lin2017,Selinger2011}, and future work in HCI could incorporate this discourse into the design and development of novel, virtue-oriented systems. Additionally, morally-relevant nudging has also started to be incorporated in public platforms outside academia---for instance Instagram experimenting with hiding like counts \cite{Perez2021} and Twitter encouraging users to read articles before retweeting links \cite{Vincent2020}---and HCI researchers can contribute by measuring the effectiveness of such efforts and discover new opportunities in this direction. 

Another possible application of prudential judgment in HCI bears mentioning: that of helping designers make better ethical decisions throughout the process. Related to this, some work in HCI has investigated how design practitioners reason through ethical dilemmas in their work. This includes some of the work mentioned in the previous section on ethical awareness \cite{Boyd2021,Gray2019b} as well as other studies engaging with moral judgment \cite{GrayChiv2019,Gray2015}. Further work in this vein could provide a foundation for educating designers on moral judgment as part of helping designers cultivate virtue.

\subsection{Extension of Moral Concern}

The final practice of cultivating virtue is extending a caring attitude beyond its initial scope. If the initial scope of our morality is to care for ourselves and our family, this practice asks us to care also for our neighbors, our fellow citizens, our fellow humans, our fellow creatures, and perhaps beyond \cite[see also][]{Floridi2013,Singer1981}. This must be done appropriately, rather than indiscriminately---to the right degree, at the right time, to the right entities and in the right way \cite{Vallor2016}. Vallor contends that this is the most powerful practice of these seven, and also the most challenging. 

An interesting parallel to the extension of moral concern is found in the extension of disciplinary concern of HCI over the course of its history. In a famous 1990 paper, Grudin described the historical trajectory of user interface design as ``reaching out'' from the computer hardware to further and vaster realms of the human social world \cite{Grudin1990}. Similarly, early visions of ubiquitous computing \cite{Weiser1991} presaged our modern day, when digital technologies are woven into the fabric of everyday life. These changes resonate with the paradigm shifts in HCI research and theory more broadly toward dynamic social contexts \cite{Harrison2011}. As well, HCI has expanded from its historical interest in making systems more usable to appealing to deeper human values, such as justice, equality, privacy, meaning and more \cite{Gurses2018,Mekler2019,Vandenhoven2015,WrightMcCarthy2010}. Perhaps inevitably, then, there is some precedent in HCI for the extension of specifically moral concern, such as in considering the needs and welfare of not only end users but a broader set of direct and indirect stakeholders, which is a core tenet of value sensitive design (VSD) \cite{Friedman2019}.

Though VSD and similar design approaches are gaining traction, there is still room for the extension of moral concern to proliferate among designers and developers---particularly as society continues to grapple with the unexpected consequences of new technologies. To this end, Vallor poses a number of questions: 

\begin{quote}
Are certain kinds of technology expanding or narrowing the scope of our moral concern, or do they exert no influence on this aspect of moral practice? Might some emerging technologies change \emph{how} we express our moral concern for others? Might they allow some forms of care, compassion, and civility to be expressed more easily than others? Could some emerging technologies exacerbate moral tribalism, neglect, or incivility, shrinking the circle of our moral concern for others? Might other technologies have the opposite effect, encouraging the exercise of moral imagination and perspective-taking that enrich our capacities for moral extension? How can we drive more resources into development of the latter, rather than the former? What is the potential risk to human flourishing if we can't, or won't? \cite[p.~117]{Vallor2016}
\end{quote}

\section{Criticisms of Virtue Ethics}

When considering any position or theory, it is indispensable to engage with arguments both for and against. As a millennia-old tradition, virtue ethics has seen a number of critiques. This section discusses three central critiques relevant to HCI---situationism, scalability and individualism---along with responses from virtue ethicists. 

\subsection{The Situationist Critique}

Virtue ethics posits that we have virtues and vices that manifest in character, and that these qualities are dynamically stable. For example, an honest person will tend to tell the truth. Starting in the 1960s, some psychologists began to cast doubt on whether people indeed have such qualities, suggesting instead that we respond much more locally to each particular situation---hence this became known as the ``situationist'' critique. For example, situationists would say there's no such thing as an honest person; a person simply lies or tells the truth in response to the pressures and opportunities of each particular situation. Situationists argue that the empirical data supports situationism and refutes both virtue ethics and personality psychology \cite[e.g.,][]{Harman1999}.

However, the empirical claims of situationists have themselves been called into question. First, these claims rely on over-generalizations that are not supported by the findings that situationists cite \cite[ch.~4]{Wright2021}. Moreover, the psychological studies that formed the basis of the situationist critique, such as those by Milgram and Zimbardo, have recently been revealed as dubious, if not fraudulent \cite{Bregman2020,Singal2021,Texier2019}. Next, more recent research demonstrates that even though there is situational variation in people's actions, there are stable qualities that emerge over longer periods of time, just as virtue ethics would predict \cite{Vallor2016,Wright2021}. Indeed, in a recent review, Wright et al. emphasize that even perfectly virtuous people need not be infallible, as humans are by definition limited (we run out of time, suffer migraines, experience major challenges, etc.) \cite[p.~17]{Wright2021}. Considering all this, Wright et al. offer a model of virtue that accounts for both the dynamism and stability of human behavior \cite{Wright2021}. 


\subsection{The Scalability Critique}

Next, critics have observed that virtue ethics emerged in antiquity, when the world was much different than it is today---societies were smaller, simpler, more closed, and more homogenous---and have suggested on these grounds that virtue ethics may no longer be actionable or useful. These critics, such as Luciano Floridi, suggest that virtue ethics doesn't \emph{scale} \cite{Floridi2013}. For instance, virtue ethics claims that morality can spread from particular individuals to families, communities and societies through the bonds of learning and caring. Regarding this claim, critics may be concerned that though learning about the virtues may help any given individual, its mechanisms are not sufficient to spread virtue among communities and societies in today's world, given the unpredictable emergent properties of complex and open societies \cite{Floridi2013}.

This critique seems to overlook, first and foremost, the reality that human culture does spread, even in today's complex and open world (perhaps especially so). Consider creative works (Harry Potter, Pok\'emon, artistic styles) and social movements (animal rights, free software, Black Lives Matter). If other cultural components can scale and spread, why should morality be uniquely unscalable? Indeed, Vallor makes an extended case that the openness and complexity of today's world are precisely why virtue ethics is the best choice for guiding technology design compared to other ethical theories \cite{Vallor2016}. 

As an outgrowth of the scalability critique, applied specifically to digital technology, we might make the observation that technology design and development today increasingly takes place at a few large companies within a corporate shareholder model, prioritizing short-term financial returns over the long-term social good. Observing this, adherents to the scalability critique of virtue ethics would suggest that cultivating virtue even among communities of consumers is not enough to ensure that emerging technologies will be conducive to virtue---that large-scale systemic change is necessary. Again, this critique may overlook the power of consumer choice in directing corporations' product development as we ``vote with our wallets'' and advocate publicly. The proliferation of organic food and electric automobiles may serve as examples where this has been successful; purchasing power among a powerful few consumers was followed by government incentive and regulation as well as further corporate innovation in an upward spiral. Moreover, business leaders such as Jacqueline Novogratz provide a model for corporations that can make positive moral changes in the world while also being competitive and profitable in the market \cite{Novogratz2020}. The rise of ``B Corporations,'' which are certified for social and environmental performance, is also relevant along these lines \cite{Kimetal2016}. 

Related to the scalability critique is a concern about relativism. Given that virtue ethics has been developed in various civilizations and epochs, some touting different virtues as central, it may be impossible (or at least unjust) to expect all the world to subscribe to one strand of virtue ethics \cite[p.~167]{Floridi2013}. In response to this critique, Charles Ess has offered the analogy to technical standards. A global internet was only possible with shared protocols and other standards, even though local variation remains; in the same way, Ess argues, a globally interconnected humanity needs shared moral norms, local variation notwithstanding \cite{Ess2006}. Still, some may wonder if this is possible. On this point, Vallor writes:

\begin{quote}
Why should we think that [the global cultivation of shared virtues] is practically possible? Consider that the alternatives are to surrender any hope for continued human flourishing, to place all our hope in an extended string of dumb cosmic luck, or to pray for a divine salvation that we can do nothing to earn. \ldots~Moreover, if Aristotle had even some success in fostering the cultivation of civic virtues, if Buddhism has encouraged any more compassion and tolerance, or Confucianism more filial care and loyalty, what is to prevent a new tradition from emerging around the technomoral virtues needed for human flourishing today? \cite[pp.~56--57]{Vallor2016}
\end{quote}

\subsection{The Individualist Critique}

Finally, because virtue ethics begins with the cultivation and activity of individuals, it may not seem appropriate as a global ethic when the most pressing moral issues are more collective than individual. Floridi suggests that the ``genuine vocation'' of virtue ethics is the individual, and it cannot readily apply to societies (but see the previous subsection); moreover, he suggests it is too Euro-centric to be appropriate for the whole world. Floridi worries that ``if misapplied,'' virtue ethics will foster narrow individualism, whereas what the world needs is broader moral cooperation and collaboration \cite[p.~167]{Floridi2013}.

First, Hans Jonas \cite{Jonas1985} has noted that this critique stands for many ethical theories developed before the 21st century; such theories were meant to address moral questions that individuals face, not collectives, and in circumstances when that person's actions would unfold within a foreseeable, stable present. More deeply, it seems that this critique is aimed at a straw-man version of virtue ethics. It is not the case that virtue ethics is narrowly concerned with a solipsistic individual, nor that it ignores systemic, situational or environmental effects \cite{MacIntyre1981}. As Vallor writes, ``virtue ethics treats persons not as atomistic individuals confronting narrowly circumscribed choices, but as beings whose actions are always informed by a particular social context of concrete roles, relationships, and responsibilities to others'' \cite[p.~33]{Vallor2016}. According to virtue ethics, we cannot cultivate virtue on our own; it takes a whole community, including parents, other family members, teachers, and friends---as well as designers, business and civic leaders, etc. Virtue ethics points to public morality just as much as to private morality, as ``a civil person will have a strong interest in supporting and maintaining her fellow citizens' capacities for excellence in public deliberation and action'' \cite[p.~143]{Vallor2016}. Critics such as Floridi ignore that virtue ethics flourished not only in ancient Greece; Confucian and Buddhist formulations of virtue ethics were always community-oriented. Even Aristotle, at the root of Western liberalism, described a virtue of civic friendship \cite{AristotleEthics}. As described in response to the scalability critique, virtue ethics is as much about infrastructure as it is about individuals' character qualities. 

\section{Discussion: Toward Virtue-Oriented Research and Design}

This paper has provided a guide to virtue ethics for HCI research and design. In particular, it detailed the seven practices of virtue cultivation as synthesized by Vallor \cite{Vallor2016}. The review of existing HCI work using virtue ethics showed several lacunae in the field's understanding of virtue ethics, and the review of HCI work pertaining to the seven virtue-building practices revealed many opportunities for further integration of virtue ethics in HCI research and design. This section reflects on those opportunities.

As we have seen, there is much existing work in HCI that relates to the seven practices of virtue cultivation, but mostly it has been developed amorally (for example, in terms of habituation rather than specifically \emph{moral} habituation). Future work in these areas can readily contribute to ethical design by orienting toward the virtues (for example, helping people build habits of honesty or courage).

According to virtue ethics, these practices can be fruitfully developed one at a time, but it may be even more powerful to engage several of them in tandem, as the practices are intertwined and mutually supporting \cite{Vallor2016}. This suggests that there are unrealized webs of connection between seemingly disparate areas of work in HCI, such as between choice architecture, the social roles of robots, and reflective informatics. The strongest virtue-oriented future work would bring together these research areas toward the pursuit of the good, both in terms of understanding existing systems and designing new ones.
 
Regarding design, Vallor's own work in \emph{Technology and the Virtues} demonstrates how virtue ethics can play an indispensable role in the design of digital technologies. The final chapters of the book present several case studies on how emerging technologies such as social media platforms and drones could be further developed toward the good (by attending to the virtues) or not (by ignoring them, or perhaps by attending to vices). As Vallor shows, such case studies may spark critical and creative thinking in designers. For example, she asks how social media platforms might cultivate a shared respect for truth beyond ``just handing everybody a louder microphone'' \cite[p.~179]{Vallor2016}, and she compels us to consider the deep emotional and educational roles of caring and being cared for that may be overlooked by creators of care robots \cite[p.~223]{Vallor2016}. Amidst these case studies, Vallor offers a conceptual methodology for ensuring designed systems are conducive to human virtue.

This points to an opportunity to further integrate existing frameworks of ethical design in HCI with virtue ethics. There already exist a number of frameworks and theoretical approaches that offer a way to consider various stakeholders' moral values in the design process. VSD, mentioned above, is perhaps the most well known among these. Other examples include Helen Nissenbaum's on values in design \cite{Nissenbaum2001} and Katie Shilton's on value conflicts \cite{Shilton2018}. The argument made in this paper would suggest that virtue ethics can supplement and perhaps further these perspectives; the work of Reijers and Gordijn on ``virtuous practice design'' provides one example to this end \cite{Reijers2019}. However, a full analysis of this question must be left for future work. 

Additionally, there are empirical opportunities for moving HCI work in a moral direction by engaging with virtue ethics. First, to determine whether and how well interactive systems help users cultivate virtue, researchers could conduct longitudinal studies using psychological measurements of virtue \cite[see][]{Wright2021} before and after a person's engagement with the system or a prototype. Next, regarding the challenge of virtuous technology within the corporate business structure discussed above, future research should explore how designers in industry engage (or not) with virtues and/or values as part of their work, and how these engagements relate to product development. To be sure, there is some existing work along these lines \cite[e.g.,][]{GrayChiv2019}, if not specifically framed with virtue ethics. Future research may explore how real-world designers respond to and apply the framework of virtue cultivation presented here, and other research may contribute to the development of specific design tools tuned to virtue development \cite[see][]{Chivukula2021}. 

\section{Conclusion}

Though the ethics of technology is becoming something of a trendy topic today, questions of morality have been with us since the dawn of computing. As Norman Cousins observed, in an essay from 1966: ``The question persists and indeed grows whether the computer makes it easier or harder for human beings to know who they really are, to identify their real problems, to respond more fully to beauty, to place adequate value on life, and to make their world safer than it now is'' \cite{Cousins1989}. That essay has been reprinted numerous times, showing its perennial relevance. 

As we continue to seek moral guidance for digital technology, this paper suggests that we not overlook virtue ethics. Though virtue ethics has been largely ignored and sometimes misunderstood within HCI, this paper has shown how it is eminently relevant and readily applicable to HCI research and design.

In closing, we might wonder why to date virtue ethics has been ignored within HCI---why so many scholars seem to be unaware of it, or immediately discard it as an option. Perhaps this is because virtue is difficult to operationalize in strictly technical terms, and in our technocratic age we tend to prefer purely technical solutions, as T. S. Eliot's lines in the epigraph of this paper suggest. For better or worse, virtue is tuned toward the human, rather than just the technical. This is a constraint of reality, if one that some would prefer to ignore. It means that, just as Light and Akama observe that technology cannot be designed ``for mindfulness'' \cite{Akama2015}, we cannot hope to design technologies that automatically create virtue. 

But this is no reason to discard virtue ethics. On the contrary, as Vallor contends, ``important as they are, the engineers of the 21st century who fashion code for machines are not as critical to the human mission as those who must fashion, test, and disseminate \emph{technomoral} code for humans---new habits and practices for living well with emerging technologies'' \cite[p.~254]{Vallor2016}. In applying virtue ethics to HCI, then, we must remember that we aren't designing ``for virtue'' but rather designing to support the human practices of cultivating virtue, from moral habituation to extending moral concern. Design to support human virtue may be exactly what is needed to start off a flywheel of morality. 

\begin{acks}
The title of this paper is an homage to Vallor's book \emph{Technology and the Virtues: A Philosophical Guide to a Future Worth Wanting} \cite{Vallor2016}, which serves as a foundation and inspiration for the present work. I am grateful to anonymous comments on a draft of this paper for guidance in improving it.
\end{acks}

\bibliographystyle{ACM-Reference-Format}
\bibliography{refs}

\end{document}